\documentstyle[prl,aps,twocolumn,graphics,epsfig]{revtex}

\begin{document}
\draft
\title{The linewidth of a non-Markovian atom laser}
\author{J.J. Hope$^{1,2,\dag}$,
G.M. Moy$^{2}$, M.J. Collett$^{1}$ and C.M. Savage$^{2}$}
\address{$^{1}$Department of Physics,
University of Auckland, New Zealand \\
$^{2}$Department of Physics and Theoretical Physics,
Australian National University,
ACT 0200, Australia.\\
$^{\dag}$ email: jjh@phy.auckland.ac.nz }
\date{\today}
\maketitle

\begin{abstract}

We present a fully quantum mechanical treatment of a single mode atom 
laser including pumping and output coupling.  By ignoring atom-atom 
interactions, we have solved this model without making the 
Born-Markov approximation.  We find substantially less gain narrowing 
than is predicted under that approximation.
\end{abstract}

\pacs{03.75.Fi,03.75-b,03.75.Be}
\narrowtext

An atom laser is a device which produces a coherent atomic de Broglie 
wave analogous to the coherent light wave produced by an optical 
laser.  Unlike optical lasers, existing experimental atom lasers are 
not pumped \cite{MITExpts,Anderson}.  Consequently they do not exhibit 
gain narrowing, the phenomenon whereby the output linewidth is much 
narrower above threshold than below threshold.

Atom laser models based on the Born-Markov approximations (BMA) 
predict gain narrowing, but they fail for a range of physically 
interesting parameter regimes \cite{Hope97a,Moy97b,Moy99}.  In the BMA 
a laser, optical or atom, is described by a quantum optical master 
equation \cite{Walls}.  In the simplest case this models an oscillator 
subject to gain and loss, due to pumping and output coupling 
respectively.  The linewidth is associated with the net 
dissipation in the system, which is the difference between the gain 
and loss.  This difference decreases as the laser is pumped above 
threshold, giving rise to the gain-narrowed laser linewidth.
This idealised Schawlow-Townes linewidth has its fundamental origin in 
spontaneous-emission-driven phase diffusion \cite{SchawlowTownes}.  By 
analogy, Graham \cite{Graham98} has estimated the ultimate 
atom laser linewidth due to scattering of thermally excited phonons.  
Previous studies based on the BMA have shown that certain pumping 
processes, such as evaporative cooling, severely broaden the atom laser 
linewidth beyond this fundamental limit \cite{Wiseman95,Wiseman96}.

In the following we present a quantum mechanical analysis of the atom 
laser which does not make the BMA. We find that the gain narrowing is 
several orders of magnitude less than predicted using the BMA. Our 
methods may be useful for other non-Markovian systems, such as 
spontaneous emission in optical band-gap materials \cite{Vats98}.

Models based on the nonlinear Schr\"{o}dinger equation are able to 
include interatomic interactions 
\cite{Naraschewski97,Steck97,Zhang98a,Kneer98}.  However they 
implicitly assume that the lasing field is in a coherent state, which 
makes it impossible to calculate the linewidth of the resulting 
output, since no information remains about the quantum statistics.  
Interatomic interactions are difficult to include in a full quantum 
mechanical model, so our model assumes that they are negligible.  This 
is an accurate description of very dilute systems.

We have previously shown that a simple model of atom laser output 
coupling produces a nondispersing state which prevents a pumped atom 
laser from reaching a steady state \cite{Moy99}. Adding 
gravity destroyed the non-dispersing state, overcoming the problem. 
Hence the atom laser modeled in this paper includes the effect of 
gravity on the output atom field.

We model the atom laser by separating it into three parts.  The lasing 
mode is an atomic cavity with large energy level separation.  We 
assume that the cavity is single mode, with annihilation 
operator $a$ and a Hamiltonian $H_{s}$.  The external atomic field 
has a different electronic state, so the atoms are no 
longer affected by the trapping potential.  We model the 
external modes with the field operator $\psi(x)$ and the 
Hamiltonian $H_{o}$.  The operators $a$ and $\psi(x)$ satisfy the 
normal boson commutation relations.  The pump reservoir is coupled to 
the cavity by an irreversible process.  At this stage, we will 
describe the pump by the Hamiltonian $H_{p}$, which also couples the 
atoms from a pump reservoir into the system mode.  The coupling 
between the lasing mode and the output modes is
\begin{equation}
        H_{int} = -i\hbar \left(\xi(t) \;a_{I}^{\dag}(t)-\xi^{\dag}(t)
        \;a_{I}(t)\right) ,
        \label{eq:Hint}
\end{equation}
where we have introduced the interaction picture operators
\begin{eqnarray}
        \nonumber
        \xi(t) &=& \int dx\; \kappa(x) \psi_{I}(x,t) , \\
        \nonumber
         \psi_{I}(x,t)& = &e^{i H_{o} (t-t_{0})/\hbar}\;\psi(x,t_{0}) \;
e^{-i H_{o} (t-t_{0})/\hbar} ,
        \label{eq:psiI}  \\
        \nonumber
        a_{I}(t) & = & e^{i/\hbar (H_{s}+H_{p})
(t-t_{0})} \;a(t_{0})\; e^{-i/\hbar (H_{s}+H_{p}) (t-t_{0})} .
        \label{eq:aI}
\end{eqnarray}
The shape of the coupling $\kappa(x)$ is determined largely by the 
spatial wavefunction of the laser mode.

Using the unitary evolution operator corresponding to the interaction 
Hamiltonian Eq.(\ref{eq:Hint}), we find in the Heisenberg
picture,
\begin{equation}
        \psi_{H}(x,t) = \psi_{I}(x,t) - \int_{t_{0}}^{t} ds\;F(x,t,s)
        a_{H}(s) ,
        \label{eq:psiH}
\end{equation}
where $\psi_{H}(x,t)$ and $a_{H}(t)$ are Heisenberg operators, and
\begin{eqnarray}
        F(x,t,s) & = & [\psi_{I}(x,t),\xi^{\dag}(s)]
        \nonumber  \\
         & = & \int dy \;\kappa^{*}(y) [\psi_{I}(x,t),\psi_{I}^{\dag}(y,s)]
        \nonumber  \\
         & = & \int dy\; \kappa^{*}(y) G(x,t,y,s)
        \label{eq:Fdef}
\end{eqnarray}
where $G(x,t,y,s)$ is the Green's function propagator due to the output 
Hamiltonian, $H_{o}$, only.  These functions can be written in closed 
form for several useful cases, including free space, free space with 
gravity, and a repulsive Gaussian potential.  We may use 
Eq.~(\ref{eq:psiH}) to calculate any observable of the output field, 
providing we know the complete history of the system, $a_{H}(s)$.

To calculate the output energy flux we transform our interaction
Hamiltonian into the basis of the energy eigenstates of the output
modes:
$H_{o} =  \int dp \;\hbar\:\omega_{p} c_{p}^{\dag} c_{p}$,
where $c_{p}$ is the annihilation operator associated with the 
eigenstate of $H_{o}$ that has a position space wavefunction 
$u_{p}(x)$ and energy $\hbar \omega_{p}$.  Defining $\bar{\kappa}(p) 
= \int dx\;u_{p}(x) \:\kappa(x)$, the output energy flux in terms of 
the two time correlation of the system is
\begin{equation}
        \frac{d\langle c_{p}^{\dag} c_{p}\rangle}{dt} = 2\; |\bar{\kappa}(p)|^{2}\;
        \Re{\left(\int_{0}^{t} ds\;e^{-i \omega_{p} \Delta t} \langle a^{\dag}(t)
        a(s)\rangle\right)},
        \label{eq:OutputSpectrum}
\end{equation}
where $\Re$ denotes the real part, and $\Delta t=t-s$.  This assumes 
that at time $t=0$, the output field was in the vacuum state.

When the output field is in free space and the only term in $H_{o}$
is the kinetic energy, then the eigenstates are the momentum
eigenstates.  In this case, $\bar{\kappa}(p)$ is just the Fourier
transform of $\kappa(x)$.  When there is a gravitational field,
the eigenstates are Airy functions with a displacement which depends
on the energy:
\begin{equation}
        u_{p}(x) = {\cal N} Ai[\beta (x - \hbar \omega_{p} / (mg) )]
        \label{eq:graveig}
\end{equation}
where $\mathcal{N}$ is a normalisation constant, the length scale 
is given by $\beta = (2 m^{2} g/\hbar^{2})^{1/3}$, and $m$ is the 
atomic mass.  In this case $\bar{\kappa}(p)$ must be calculated 
numerically.

Following Scully and Lamb we model pumping by the injection of a 
Poissonian sequence of excited atoms into the atom laser 
\cite{Walls,Scully67}.  These atoms may spontaneously emit a photon 
and make a transition either into the atom lasing mode or into other 
modes of the lasing cavity.  For simplicity, we consider a 
two-mode approximation.  To obtain the pumping term, we consider the 
effect of a single atom injected into the atom laser, and then extend 
this to describe the effect of a distribution of atoms.  This gives 
the master equation \cite{Wiseman97}
\begin{eqnarray}
\left(\dot{\rho}\right)_{\mbox{pump}} &=& r {\cal D} [a^{\dag}] \left(
n_{s} + {\cal A}[a^{\dag}]\right)^{-1}
\rho ,
\label{Eq.MasterEqn3}
\end{eqnarray}
where $r$ is the rate at which atoms are injected into the cavity, and 
$n_{s}$ the saturation number. The superoperators ${\cal D}$ 
and ${\cal A}$ are defined by
\begin{eqnarray}
{\cal D}[c] &=& c \rho c^{\dag} - {\cal A}[c], \\
{\cal A}[c] \rho &=& \frac{1}{2}(c^{\dag}c \rho + \rho c^{\dag}c).
\label{eq:SuperOps}
\end{eqnarray}

In our
particular model $n_{s}$ depends on the ratio of the probability that
an atom will spontaneously emit into the lasing mode to the
probability that the atom will emit into another mode.

If we make the Born approximation, trace over the output modes, and
then make the Markov approximation, we
can write the damping term of the master equation as \cite{Moy99}
\begin{eqnarray}
\left(\dot{\sigma}\right)_{\mbox{damp}} &=& \gamma_{BM} {\cal D}[a]
\sigma ,
 \label{eq:MarkovDamping}
\end{eqnarray}
where $\gamma_{BM}$ is related to the above threshold mean atom 
number $\bar{n}$ and saturation photon number $n_{s}$ by
\begin{equation}
\gamma_{BM} = r / (\bar{n}+n_{s}) .
 \label{eq:gammabm}
\end{equation}
The full equation of motion is then
\begin{eqnarray}
\frac{d \langle a^{\dag}\rangle}{dt} &=&
\left( P - \frac{\gamma_{BM}}{2}\right)
\langle a^{\dag}\rangle
\approx -\frac{r}{4 \bar{n}^{2}} \langle a^{\dag}\rangle ,
  \label{eq:dadtMarkov}\\
P &\approx& r / \{ 2(\bar{n}+n_{s})+1 \} ,
\end{eqnarray}
which has an error term proportional to $r/\bar{n}^{3}$ if we assume 
that the system is close to a coherent state.
The solution to this equation is an exponential decay, and the energy
spectrum is therefore Lorentzian, with a width of
\begin{equation}
\Gamma_{BM} = r/(4 \bar{n}^{2}) .
\label{eq:BMlinewidth}
\end{equation}
If we do not make the Born or Markov approximations, but we do assume
that the trap population is localised around some (at this stage
unknown) value $\bar{n}$, we find
\begin{eqnarray}
        \frac{\partial}{\partial t}\ \langle a^{\dag}(t) \rangle  =
         (i \omega_{o} + P)\langle
            a^{\dag}(t)\rangle  + \int dx \;
            \kappa^{*}(x) \langle \psi^{\dag}_{x,t}\rangle
        \label{eq:Dadag}
\end{eqnarray}
well above threshold.  Note that we can no longer relate $\bar{n}$ 
directly to the physical parameters of the problem using 
Eq.(\ref{eq:gammabm}), which has used the Born approximation.  Since 
$P$ determines $n_{s}$, we seek an iterative method to produce a 
self-consistent solution.

Under our assumptions the pumping is effectively linear, so we may use 
the quantum regression theorem.  Using Eq.(\ref{eq:psiH}), we may 
derive the following Volterra convolution type integro-differential 
equation for the two time correlation function:
\begin{eqnarray}
        \nonumber
        \frac{\partial}{\partial \tau} &\langle a^{\dag}(t+\tau) a(t)
        \rangle = (i \omega_{o} + P) \langle a^{\dag}(t+\tau)
        a(t)\rangle \\
        &\:\:\:\;\;\;- \int_{0}^{t+\tau}
        du \;f^{*}(t+\tau-u) \langle a^{\dag}(u) a(t)\rangle,
        \label{eq:TTCEOM}
\end{eqnarray}
where $\tau > 0$, and $f(\Delta t)$ is the memory function
\begin{equation}
        f(\Delta t) = \int dx \; \kappa(x) F(x,\Delta t).
        \label{eq:fdef}
\end{equation}
We have assumed that $F(x,t,s)$ is a function of $\Delta t=t-s$.
This again assumes that at time $t=0$, the output field was in the 
vacuum state.  This equation is not sufficient to specify the dynamics 
of the cavity, as it is only a single partial integro-differential 
equation in a two dimensional space.  We also require the 
integro-differential equation for the intracavity number, which we can 
generate in a similar manner.  Well above threshold, we obtain
\begin{equation}
        \label{eq:NEOM}
        \frac{d}{d t}\langle a^{\dag}a\rangle (t) =
        r \\
        \nonumber
        - \int_{0}^{t}
        ds \;2 \Re\{f^{*}(\Delta t) \langle a^{\dag}(s) a(t)\rangle\}.
\end{equation}

These equations are difficult to solve in general, but can be solved 
in various limits.  For example, if the kernel $f(t)$ is a 
$\delta$-function, as in the broadband limit of the optical laser, 
then the equations would become local, and the solution is an 
exponential.  Although the broadband limit can be a good approximation 
for the atomic case as well \cite{Moy97b}, in general the atoms will 
disperse, which gives the system an irreducible memory.

In order to produce analytical forms for the memory functions we 
assume that the coupling is Gaussian, and that there is no net 
momentum kick given to the atoms:
\begin{equation}
        \kappa(x) = \sqrt{\gamma} \left(2
        \sigma_{k}^{2} / \pi \right)^{1/4} \exp[-(\sigma_{k}\;x)^{2}],
        \label{eq:kappadef}
\end{equation}
where $\hbar \sigma_{k}$ is the momentum width of the coupling and
$\gamma$ is the strength of the coupling.
In the presence of a gravitational field, $V=mgx$, the Green's function
in Eq.(\ref{eq:Fdef}) can be found as a standard result
\cite{Jack99a,Schulman81}:
\begin{eqnarray}
        \label{eq:Greensfn}
        &&G(x,t,y,s)=\sqrt{\frac{1}{4\pi i \lambda \Delta t}} \times \\
        \nonumber
        &&\;\;\;\exp[({\frac{i(x-y)^{2}}{4 \lambda \Delta t} -\frac{ig\Delta t(x+y)}{4
        \lambda}- \frac{ig^{2}\Delta t^{3}}{48 \lambda}})],
\end{eqnarray}
where $\lambda = \hbar/(2m)$, and $\Delta t = t-s$.  This leads to the 
following form for the memory function,
\begin{equation}
        f(\Delta t)  =  \frac{\gamma \;\;\exp({-\frac{g^{2}\: \Delta
        t^{2}}{32\:\lambda^{2}\:\sigma_{k}^{2}}})
        \;\;\exp({\frac{-i\:g^{2}\:\Delta t^{3}}{48
         \:\lambda}})}{\sqrt{1+ 2\:i\:\lambda\:\Delta
        t\:\sigma_{k}^{2}}}.
        \label{eq:fgrav}
\end{equation}
For coupling into free space, $g=0$, and $f(\Delta t)$ goes as 
$1/\sqrt{\Delta t}$ in the long time limit.  There are no useful 
approximations when $g\neq 0$.  This is because the broadband limit of 
the integrals involving $f(\Delta t)$ become unbounded in amplitude, 
and their convergence is due to their highly oscillatory nature.

Eq.~(\ref{eq:TTCEOM}) and Eq.~(\ref{eq:NEOM}) do not form a
standard pair of partial integro-differential equations.  The
derivative in Eq.~(\ref{eq:TTCEOM}) is only defined for $\tau>0$, and
so we cannot require that the solution obey this equation through the
whole domain of the integral.  This means that Eq.~(\ref{eq:TTCEOM})
cannot be integrated to find the solution, as we do not actually know
the derivative at any point.  Since the two-time correlation is
Hermitian, we can we rewrite the integral so that the domain 
remains in the $\tau>0$ plane, but we still do not have a continuously
defined derivative along the length of the integral.

We proceed by making an ansatz which uses the solution of 
Eq.~(\ref{eq:TTCEOM}) which has been extended into the region $\tau<
0$.  We then use the $\tau>0$ portion of this solution to substitute
into the two time correlation in Eq.~(\ref{eq:NEOM}).  We introduce
the function $J(t)$, which is the solution of the equation
\begin{equation}
        \frac{d J(t)}{d t} = (i \omega_{o}+P) J(t) -
        \int_{0}^{t} ds\;f^{*}(\Delta t) J(s).
        \label{eq:Jdef}
\end{equation}

This means that
\begin{equation}
        \langle a^{\dag}(t+\tau) a(t) \rangle = \langle a^{\dag} a \rangle
        (t) J(t+\tau)/ J(t)
        \label{eq:S1}
\end{equation}
is a solution of Eq.~(\ref{eq:TTCEOM}) with the correct initial
condition at $\tau=0$.  We then substitute this result into
Eq.(\ref{eq:NEOM}):
\begin{equation}
        \frac{d\langle a^{\dag} a \rangle (t)}{dt} = r - \int_{0}^{t}
        ds\;\langle a^{\dag} a \rangle (s)\;\Re{\left(\frac{2 f(\Delta t)
        J(t)}{J(s)}\right)}.
        \label{eq:S2}
\end{equation}

Solving these two equations gives the two time correlation for the
lasing mode, from which we may find the properties of the output
field.  It is only consistent with our linearisation of the pumping if
the number of atoms in the trap, $\bar{n}$, converges to the value
which originally produced the parameter $P$.  Since we require $P$ to
generate the solution, and $\bar{n}$ is simply the long time limit of
Eq.~(\ref{eq:S2}), the effective free parameter is $n_{s}$.  This
threshold parameter must be much smaller than $\bar{n}$, so we search
for a value of $P$ which gives the result $\bar{n}\gg n_{s}$.

Once it is established that Eq.~(\ref{eq:S2}) is approaching a stable
steady state, a fast way of finding it is to set the
derivative to zero, and assume that $\langle a^{\dag} a \rangle
(s)=\bar{n}$ over the support of the kernel.  This gives
\begin{equation}
        \bar{n} = r \left(
        \int_{0}^{t} ds\;\Re \left(\frac{2 f(\Delta t) 
        J(t)}{J(s)} \right) 
        \right)^{-1} .
        \label{eq:nbarquick}
\end{equation}

We transform Eq.~(\ref{eq:Jdef}) into a rotating frame by introducing 
the function $N(t) = J(t)\;\exp(-i \omega_{o} t - P t)$ and rewriting 
it in terms of $N(t)$:
\begin{equation}
        \frac{d N(t)}{d t} = -
        \int_{0}^{t} ds\;H(\Delta t) N(s)
        \label{eq:otherNeom}
\end{equation}
where $H(\Delta t) = f^{*}(\Delta t) \exp(-[i \omega_{o} + P]\Delta t)$.  An 
effective numerical method for solving this equation is direct 
integration using a second-order algorithm for both the integration 
and the calculation of the integral to find the derivative at each 
timestep \cite{NumericalRecipes}.

We now calculate the output properties of an atom laser. We use the trap 
frequency $\omega_{o}=2 \pi \times 123$ Hz \cite{MITExpts}, an atomic 
mass of $m=5 \times 10^{-26}$kg, a gravitational acceleration of $g=9.8 
\sin(0.18)$ m s$^{-2}$, and the coupling given by 
Eq.~(\ref{eq:kappadef}) with a momentum width $\sigma_{k} = 4.4 \times 
10^{5}$ m$^{-1}$.  We use a damping constant of $\gamma=2.0 \times 
10^{4} s^{-2}$, and a threshold of $n_{s}=47$. 
As the pumping rate $r$ increases, the steady state number of atoms 
in the cavity $\bar{n}$ increases, and the modulus of the two time correlation 
decays more slowly.  The energy spectrum of the output flux is 
proportional to the Fourier transform of the two time correlation 
through Eq.~(\ref{eq:OutputSpectrum}), so the laser shows gain 
narrowing.

In Table~1 we show the results of these calculations.  The linewidth 
of the output flux, $\Gamma$, is calculated directly from the 
two time correlation using Eq.~(\ref{eq:OutputSpectrum}).  This is 
compared to the linewidth given by the Born-Markov approximation 
$\Gamma_{BM}$, Eq.~(\ref{eq:BMlinewidth}).

\begin{figure}
\begin{center}
\epsfxsize=\columnwidth
\epsfbox{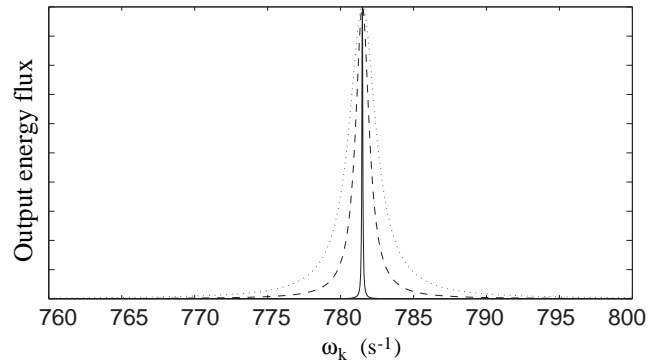}
\end{center}
\caption{The spectra of the output energy flux for three different 
pumping rates.  The dotted, dashed and solid lines represent $r=40 
s^{-1}$, $r=80 s^{-1}$ and $r=800 s^{-1}$ respectively.  Other 
parameters are given in the text.}
\label{fig:linenarrowing}
\end{figure}

The atom laser linewidth predicted by our theory is several orders of 
magnitude larger than that predicted using the BMA result.  We plot 
the spectral flux corresponding to three of these pumping rates in 
Fig.~\ref{fig:linenarrowing}.  The vertical scale is normalised to the 
peak height for each plot so that the width of the spectra can be 
easily compared.  The spectra are almost Lorenztian, but have a 
drastically different width and are slightly shifted compared to the 
results under the BMA.

Our model does not include atom-atom interactions, and therefore only 
works when the atomic field is very dilute.  The advantage of our 
model is that, unlike mean field models, we have avoided making the 
approximation that the lasing mode is perfectly coherent.  Future work 
will involve generalising this atom laser model to include atom-atom 
interactions.

This work was supported by the Australian Research Council.  J.H.
would like to thank M.Jack, T.Ralph, H.Wiseman and M.Naraschewski for
their helpful discussions.

\begin{table}
        \centering
        \caption{Linewidths and cavity atom number as a function of pumping 
        rate $r$.}
\begin{tabular}{|c||c|c|c|}
        \hline
        $r$ ($10^{3}$/s) & $\bar{n}$ & $\Gamma$($s^{-1}$) &
        $\Gamma_{BM}$($s^{-1}$)  \\
        \hline \hline
        20 & 450 & 2.1 & 0.025  \\
        \hline
        40 & 910 & 1.1 & 0.012   \\
        \hline
        80 & 1800 & 0.56 & 0.0062   \\
        \hline
        800 & $1.8 \times 10^{4}$ & 0.035 & 0.00062  \\
        \hline
\end{tabular}
        \label{tbl:linenarrowing}
\end{table}

\end{document}